\documentclass[12pt,a4paper]{article}

\usepackage{epsfig}
\usepackage{axodraw}

\usepackage{bold-extra}

\newcommand{\mrm}[1]{\mathrm{#1}}

\newcommand{\tsc}[1]{\textsc{#1}}

\newlength{\tmplen}

\def\lsim{\mathrel{\rlap{\lower4pt\hbox{\hskip1pt$\sim$}}
    \raise1pt\hbox{$<$}}}                
\def\gsim{\mathrel{\rlap{\lower4pt\hbox{\hskip1pt$\sim$}}
    \raise1pt\hbox{$>$}}}                
 
\newcommand{\pT}{\ensuremath{p_{\perp}}}
\newcommand{\kT}{\ensuremath{k_{\perp}}}

\newcommand{\GeV}{\ensuremath{\!\ \mathrm{GeV}}}
\newcommand{\TeV}{\ensuremath{\!\ \mathrm{TeV}}}

\renewcommand{\d}{\mathrm{d}}

\newcommand{\g}{\mathrm{g}}

\newcommand{\p}{\mathrm{p}}
\newcommand{\q}{\mathrm{q}}

\newcommand{\Z}{\mathrm{Z}}

\newcommand{\qbar}{\overline{\mathrm{q}}}

\newenvironment{Itemize}{\begin{list}{$\bullet$}%
{\setlength{\topsep}{0.2mm}\setlength{\partopsep}{0.2mm}%
\setlength{\itemsep}{0.2mm}\setlength{\parsep}{0.2mm}}}%
{\end{list}}
\newcounter{enumct}

\newlength{\abstwidth}
\newlength{\captivewidth}
\begin{document}
\sloppy
 
\pagestyle{empty}
 
\begin{flushright}
LU TP 08-21\\
MCnet/08/15
\end{flushright}
 
\vspace{\fill}

\begin{center}
\renewcommand{\thefootnote}{\fnsymbol{footnote}} 
{\LARGE\bf Multiple Interactions in \tsc{Pythia 8}%
\footnote{ To appear in the Proceedings of the First International Workshop
on Multiple Partonic Interactions at the LHC (MPI@LHC), Perugia, Italy,
27-31 October 2008}%
\footnote{Work supported by the Marie Curie Early Stage Training program
``HEP-EST'' (contract number \makebox{MEST-CT-2005-019626}) and in part by
the Marie Curie RTN ``MCnet'' (contract number
\makebox{MRTN-CT-2006-035606})}}\\[10mm]
\renewcommand{\thefootnote}{\arabic{footnote}}
\setcounter{footnote}{0}
{\Large R.~Corke\footnote{richard.corke@thep.lu.se; work done in
collaboration with T.~Sj\"ostrand and in part with F.~Bechtel}} \\[3mm]
{\it Department of Theoretical Physics,}\\[1mm]
{\it Lund University,}\\[1mm]
{\it S\"olvegatan 14A,}\\[1mm]
{\it S-223 62 Lund, Sweden}
\end{center}
 
\vspace{\fill}
 
\begin{center}
{\bf Abstract}\\[2ex]
\begin{minipage}{\abstwidth}
Modelling multiple partonic interactions in hadronic events is vital for
understanding minimum-bias physics, as well as the underlying event of hard
processes. A brief overview of the current \tsc{Pythia 8} multiple
interactions (MI) model is given, before looking at two additional effects
which can be included in the MI framework. With rescattering, a previously
scattered parton is allowed to take part in another subsequent scattering,
while with enhanced screening, the effects of varying initial-state
fluctuations are modelled.
\end{minipage}
\end{center}
 
\vspace{\fill}
 
\clearpage
\pagestyle{plain}
\setcounter{page}{1}

\section{Introduction}
The run-up to the start of the LHC has led to a greatly increased interest
in the physics of multiple parton interactions in hadronic collisions.
Existing models are used to try to get an insight into what can be expected
at new experiments, extrapolating fits to Tevatron and other data to LHC
energies \cite{MPILHC}. Such extrapolations, however, come with a high
level of uncertainty; within many models are parameters which scale with an
uncertain energy dependence. There is, therefore, also the exciting
prospect of new data, with which to further constrain and improve models.

In terms of theoretical understanding, MI is one of the least well
understood areas. While current models, after tuning, are able to describe
many distributions very well, there are still many others which are not
fully described. This is a clear sign that new physical effects need to be
modelled and it is therefore not enough to ``sit still'' while waiting for
new data. It is with this in mind that we look at two new ideas in the
context of MI and their potential effects.

With rescattering, an already scattered parton is able to undergo another
subsequent scattering. Although, in general, such rescatterings may be
relatively soft, even when compared to normal $2 \rightarrow 2$ MI
scatterings, they can lead to non-trivial colour flows which change the
structure of events. Another idea is to consider partonic fluctuations in
the incoming hadrons before collision. In such a picture, it is possible to
get varying amounts of colour screening on an event-by-event basis. The
question then is, what effects such new ideas would have on multiple
interactions and how can they be included in the \tsc{Pythia}
framework?

In Section~\ref{sec:MI}, a brief introduction to the existing MI model in
\tsc{Pythia 8} is given. For more comprehensive details about what is
contained in the model, readers are directed to \cite{Sjostrand:2006za} and
the references therein. In Sections~\ref{sec:rescattering}~and~\ref{sec:ES},
an initial look at rescattering and enhanced screening is given. A summary
and outlook is given in Section~\ref{sec:conclusions}.

\section{Multiple Interactions in \tsc{Pythia 8}}
\label{sec:MI}

The MI model in \tsc{Pythia 8} \cite{Sjostrand:2007gs} is a model for
non-diffractive events. It is an evolution of the model introduced in
\tsc{Pythia 6.3} \cite{Sjostrand:2006za}, which in turn is based on the
model developed in earlier versions of \tsc{Pythia}. The earliest model
\cite{Sjostrand:1987su} was built around the virtuality-ordered parton
showers available at the time and introduced many key features which are
still present in the later models, such as $\pT$ ordering, perturbative QCD
cross sections dampened at small $\pT$, a variable impact parameter, PDF
rescaling, and colour reconnection.

The next-generation model \cite{Sjostrand:2004ef, Sjostrand:2004pf} was
developed after the introduction of transverse-momentum-ordered showers,
opening the way to have a common $\pT$ evolution scale for initial-state
radiation (ISR), final-state radiation (FSR) and MI emissions. The second
key ingredient was the addition of junction fragmentation to the Lund
String hadronisation model, allowing the handling of arbitrarily
complicated beam remnants. This permitted the MI framework to be updated
to include a more complete set of QCD $2 \rightarrow 2$ processes, with the
inclusion of flavour effects in the PDF rescaling.

The \tsc{Pythia 8} MI framework also contains additional new features
which are not found in previous versions, such as
\begin{Itemize}
 \item a richer mix of underlying-event processes ($\gamma$, J/$\psi$,
Drell-Yan, etc.),
 \item the possibility to select two hard interactions in the same event, and
 \item the possibility to use one PDF set for hard processes and another for
other subsequent interactions.
\end{Itemize}

\subsection{Interleaved $\pT$ Ordering}
Starting in \tsc{Pythia 6.3}, ISR and MI were interleaved with a common
$\pT$ evolution scale. In \tsc{Pythia 8}, this is taken a step further,
with FSR now also fully interleaved. The overall probability for the $i^{th}$
interaction or shower branching to take place at $\pT = {\pT}_i$ is given by
\begin{eqnarray}
 \frac{\d \mathcal{P}}{\d \pT} & = &
 \left( \frac{\d \mathcal{P}_{\mrm{MI}}}{\d \pT} +
 \sum \frac{\d \mathcal{P}_{\mrm{ISR}}}{\d \pT} +
 \sum \frac{\d \mathcal{P}_{\mrm{FSR}}}{\d \pT}
 \right) \nonumber \\
 & \times & \exp \left( - \int_{\pT}^{p_{\perp i-1}} \left(
 \frac{\d \mathcal{P}_{\mrm{MI}}}{\d \pT'} +
 \sum \frac{\d \mathcal{P}_{\mrm{ISR}}}{\d \pT'} +
 \sum \frac{\d \mathcal{P}_{\mrm{FSR}}}{\d \pT'}
 \right) \d \pT' \right)
 ,
 \label{eqn:pTevol}
\end{eqnarray}
with contributions from MI, ISR and FSR unitarised by a Sudakov-like
exponential factor.

If we now focus on just the MI contribution, the probability for an
interaction is given by
\begin{equation}
 \frac{\d \mathcal{P}}{\d {\pT}_i} =
 \frac{1}{\sigma_{nd}} \frac{\d \sigma}{\d \pT} \;
 \exp \left( - \int_{\pT}^{p_{\perp i-1}}
  \frac{1}{\sigma_{nd}} \frac{\d \sigma}{\d \pT'} \d \pT' \right)
 ,
 \label{eqn:MIevol}
\end{equation}
where $\d \sigma / \d \pT$ is given by the perturbative QCD $2\rightarrow2$
cross section. This cross section is dominated by $t$-channel gluon
exchange, and diverges roughly as $\d \pT^2 / \pT^4$. To avoid this
divergence, the idea of colour screening is introduced. The concept of a
perturbative cross section is based on the assumption of free incoming
states, which is not the case when partons are confined in colour-singlet
hadrons. One therefore expects a colour charge to be screened by the
presence of nearby anti-charges; that is, if the typical charge separation
is $d$, gluons with a transverse wavelength $\sim 1 / \pT > d$ are no
longer able to resolve charges individually, leading to a reduced effective
coupling. This is introduced by reweighting the interaction cross section
such that it is regularised according to
\begin{equation}
 \frac{\d \hat{\sigma}}{\d \pT^2} \propto
 \frac{\alpha_S^2(\pT^2)}{\pT^4} \rightarrow
 \frac{\alpha_S^2({\pT^2}_0 + \pT^2)}{({\pT^2}_0 + \pT^2)^2}
 ,
 \label{eqn:pt0}
\end{equation}
where ${\pT}_0$ (related to $1 / d$ above) is now a free parameter in the
model.

\subsection{Impact Parameter}
Up to this point, all parton-parton interactions have been assumed to be
independent, such that the probability to have $n$ interactions in an event,
$\mathcal{P}_n$, is given by Poissonian statistics. This picture is now
changed, first by requiring that there is at least one interaction, such
that we have a physical event, and second by including an impact parameter,
$b$. The default matter distribution in \tsc{Pythia} is a double
Gaussian
\begin{equation}
 \rho(r) \propto
 \frac{1 - \beta}{a_1^3} \exp \left( - \frac{r^2}{a_1^2} \right) +
 \frac{\beta}{a_2^3}     \exp \left( - \frac{r^2}{a_2^2} \right)
 ,
\end{equation}
such that a fraction $\beta$ of the matter is contained in a radius $a_2$,
which in turn is embedded in a radius $a_1$ containing the rest of the
matter. The time-integrated overlap of the incoming hadrons during
collision is given by
\begin{equation}
 \mathcal{O}(b) =
 \int \d t \int \d^3 x \;
 \rho(x, y, z) \;
 \rho(x + b, y, z + t)
 ,
\end{equation}
after a suitable scale transformation to compensate for the boosted nature
of the incoming hadrons.

Such an impact parameter picture has central collisions being generally
more active, with an average activity at a given impact parameter being
proportional to the overlap, $\mathcal{O}(b)$. While requiring at least one
interaction results in $\mathcal{P}_n$ being narrower than Poissonian, when
the impact parameter dependence is added, the overall effect is that
$\mathcal{P}_n$ is broader than Poissonian. The addition of an impact
parameter also leads to a good description of the ``Pedestal Effect'',
where events with a hard scale have a tendency to have more underlying
activity; this is as central collisions have a higher chance both of
a hard interaction and of more underlying activity. This centrality effect
naturally saturates at ${\pT}_{hard} \sim 10 \GeV$.

\subsection{PDF Rescaling}
In the original model, PDFs were rescaled only such that overall momentum
was conserved. This was done by evaluating PDFs at a modified $x$ value
\begin{equation}
 x_i' = \frac{x_i}{1 - \sum^{i-1}_{j=1} x_j}, 
\end{equation}
where the subscript i refers to the current interaction and the sum runs
over all previous interactions. The original model was affected by a
technical limitation in fragmentation; it was only possible to take one
valence quark from an incoming hadron. This meant that the MI framework was
limited to $\q\qbar$ and $\g\g$ final states and that it was not possible
to have ISR from secondary scatterings. By introducing junction
fragmentation, where a central junction is connected to three quarks and
carries baryon number, these limitations were removed. This allowed the
next-generation model to include a more complete set of MI processes and
flavour effects in PDF rescaling.

ISR, FSR and MI can all lead to changes in the incoming PDFs. In the case
of FSR, a colour dipole can stretch from a radiating parton to a beam
remnant, leading to (a modest amount of) momentum shuffling between the
beam and the parton.  Both ISR and MI can result in large $x$ values being
taken from the beams, as well as leading to flavour changes in the PDFs. If
a valence quark is taken from one of the incoming hadrons, the valence PDF
is rescaled to the remaining number. If, instead, a sea quark ($\mrm{q_s}$)
is taken from a hadron, an anti-sea companion quark ($\mrm{q_c}$) is left
behind. The $x$ distribution for this companion quark is generated from a
perturbative ansatz, where the sea/anti-sea quarks are assumed to have come
from a gluon splitting, $\mrm{g \rightarrow q_s q_c}$. Subsequent
perturbative evolution of the $\mrm{q_c}$ distribution is neglected.
Finally, there is the issue of overall momentum conservation. If a valence
quark is removed from a PDF, momentum must be put back in, while if a
companion quark is added, momentum must be taken from the PDF.  This is
done by allowing the normalisation of the sea and gluon PDFs to fluctuate
such that overall momentum is conserved.

\subsection{Beam Remnants, Primordial $\kT$ and Colour Reconnection}
\label{sec:BR}
When the $\pT$ evolution has come to an end, the beam remnant will consist
of the remaining valence content of the incoming hadrons as well as any
companion quarks. These remnants must carry the remaining fraction of
longitudinal momentum. \tsc{Pythia} will pick $x$ values for each
component of the beam remnants, according to distributions such that the
valence content is ``harder'' and will carry away more momentum. In the
rare case that there is no remaining quark content in a beam, a gluon is
assigned to take all the remaining momentum.

The event is then modified to add primordial $\kT$. Partons are expected to
have a non-zero $\kT$ value just from Fermi motion within the incoming
hadrons. A rough estimate based on the size of the proton gives a value of
$\sim 0.3 \GeV$, but when comparing to data, for instance the $\pT$
distribution of $\Z^0$ at CDF, a value of $\sim 2 \GeV$ appears to
be needed. The current solution is to decide a \kT value for each
initiator parton taken from a hadron based on a Gaussian whose width is
generated according to an interpolation
\begin{equation}
 \sigma(Q) = \mrm{max}
 \left( \sigma_{min},
 \sigma_{\infty} ~ \frac{1}{1+Q_{\frac{1}{2}} / Q} \right)
 ,
\end{equation}
where $Q$ is the hardness of a sub-collision, $\sigma_{min}$ is a minimal
value ($\sim 0.3 \GeV$), $\sigma_{\infty}$ is a maximal value that is
approached asymptotically and $Q_{\frac{1}{2}}$ is the $Q$ value at which
$\sigma(Q)$ is equal to half $\sigma_{\infty}$. The recoil is shared among
all initiator and remnant partons from the incoming hadrons, and the $\kT$
given to all daughter partons through a Lorentz boost.

The final step is colour reconnection. In the old MI framework,
Rick Field found a good agreement to CDF data if 90\% of additional
interactions produced two gluons with ``nearest neighbour'' colour
connections \cite{Field:2002vt}. In \tsc{Pythia 8}, with its more
general MI framework, colour reconnection is performed by giving each
system a probability to reconnect with a harder system
\begin{equation}
 \mathcal{P} = \frac{{\pT}_{Rec}^2}{({\pT}_{Rec}^2 + \pT^2)}, ~~~~~
 ~~~~~ {\pT}_{Rec} = RR * {\pT}_{0},
 \label{eqn:crec}
\end{equation}
where $RR$, ReconnectRange, is a user-tunable parameter and ${\pT}_{0}$
is the same parameter as in eq.~(\ref{eqn:pt0}). The idea of colour
reconnection can be motivated by noting that MI leads to many colour strings
that will overlap in physical space. Moving from the limit of
$N_C \rightarrow \infty$ to $N_C = 3$, it is perhaps not unreasonable to
consider these strings to be connected differently due to a coincidence of
colour, so as to reduce the total string length and thereby the potential
energy. With the above probability for reconnection, it is easier to
reconnect low $\pT$ systems, which can be viewed as them having a larger
spatial extent such that they are more likely to overlap with other colour
strings. Currently, however, given the lack of a firm theoretical basis,
the need for colour reconnection has only been established within the
context of specific models.

\section{Rescattering}
\label{sec:rescattering}

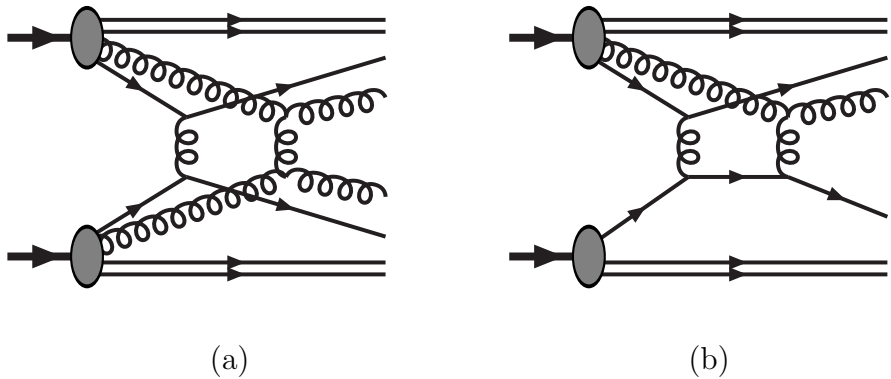
\begin{figure}
 \begin{center}
 \begin{picture}(157.5,112.5)(0,0)
  \SetScale{0.75}
  \SetWidth{4}
  \ArrowLine(10,130)(50,130)
  \ArrowLine(10,20)(50,20)
  \SetWidth{2}
  \ArrowLine(50,121)(100,90)
  \ArrowLine(50,29)(100,60)
  \Gluon(100,60)(100,90){5}{2}
  \ArrowLine(100,90)(200,120)
  \ArrowLine(100,60)(200,30)
  \Gluon(50,127)(150,90){5}{10}
  \Gluon(50,23)(150,60){5}{10}
  \Gluon(150,60)(150,90){5}{2}
  \Gluon(150,90)(200,100){5}{4}
  \Gluon(150,60)(200,50){5}{4}
  \ArrowLine(50,133)(200,133)
  \ArrowLine(50,139)(200,139)
  \ArrowLine(50,17)(200,17)
  \ArrowLine(50,11)(200,11)
  \GOval(50,130)(15,8)(0){0.5}
  \GOval(50,20)(15,8)(0){0.5}
 \end{picture}\hspace{10mm}
 \begin{picture}(157.5,112.5)(0,0)
  \SetScale{0.75}
  \SetWidth{4}
  \ArrowLine(10,130)(50,130)
  \ArrowLine(10,20)(50,20)
  \SetWidth{2}
  \ArrowLine(50,121)(100,90)
  \ArrowLine(50,25)(100,60)
  \Gluon(100,60)(100,90){5}{2}
  \ArrowLine(100,90)(200,120)
  \ArrowLine(100,60)(150,60)
  \Gluon(50,127)(150,90){5}{10}
  \Gluon(150,60)(150,90){5}{2}
  \Gluon(150,90)(200,100){5}{4}
  \ArrowLine(150,60)(200,40)
  \ArrowLine(50,133)(200,133)
  \ArrowLine(50,139)(200,139)
  \ArrowLine(50,17)(200,17)
  \ArrowLine(50,11)(200,11)
  \GOval(50,130)(15,8)(0){0.5}
  \GOval(50,20)(15,8)(0){0.5}
 \end{picture}
\end{center}

\hspace{50mm}(a)
\hspace{57mm}(b)

\caption{(a) Two $2 \rightarrow 2$ scatterings, (b) a $2 \rightarrow 2$
scattering followed by a rescattering}
\label{fig:res}
\end{figure}

A process with a rescattering occurs when an outgoing state from one
scattering is allowed to become the incoming state in another scattering.
This is illustrated schematically in Figure~\ref{fig:res}, where (a) shows
two independent $2 \rightarrow 2$ processes while (b) shows a rescattering
process. An estimate for the size of such rescattering effects is given by
Paver and Treleani \cite{Paver:1983hi}, where a factorised form is used for
the double parton distribution, giving the probability of finding two
partons of given $x$ values inside an incoming hadron. Their results show
that, at Tevatron energies, rescattering is expected to be a small effect
when compared against the more dominant case of multiple disconnected
scatterings.

If we accept MI as real, however, then we should also allow rescatterings
to take place. They would show up in the collective effects of MI,
manifesting themselves as changes to multiplicity, $\pT$ and other
distributions. After a retuning of ${\pT}_0$ and other model parameters, it
is likely that their impact is significantly reduced, so we should
therefore ask whether there are more direct ways in which rescattering may
show up. Is there perhaps a region of low $\pT$ jets, where an event is not
dominated by ISR/FSR, where this extra source of three-jet topologies will
be visible?  A further consideration is that such rescatterings will
generate more $\pT$ in the perturbative region, which may overall mean it
is possible to reduce the amount of primordial $\kT$ and colour
reconnections necessary to match data, as discussed in
Section~\ref{sec:BR}.

\subsection{Rescattering in \tsc{Pythia 8}}
If we begin with the typical case of small-angle $t$-channel gluon
scattering, we can imagine that a combination of a scattered parton and a
hadron remnant will closely match one of the incoming hadrons.
In such a picture, we can write the complete PDF for a hadron as
\begin{equation}
 f(x,Q^2) \rightarrow
 f_{rescaled}(x,Q^2) + \sum_n \delta(x - x_n)
 = f_u(x,Q^2) + f_{\delta}(x, Q^2)
 ,
\end{equation}
where the subscript u/$\delta$ is the unscattered/scattered component. That
is, each time a scattering occurs, one parton is fixed to a specific $x_n$
value, while the remainder is still a continuous probability distribution.
In such a picture, the momentum sum should still approximately obey
\begin{equation}
  \int_0^1 x \left[ f_{rescaled}(x, Q^2) + \sum_n \delta(x - x_n) \right]
  \d x = 1
  .
\end{equation}

Of course, in general, it is not possible to uniquely identify a scattered
parton with one hadron, so an approximate prescription must be used
instead, such as rapidity based. If we consider the original MI probability
given in eqs.~(\ref{eqn:pTevol})~and~(\ref{eqn:MIevol}), we can now
generalise this to include the effects of rescattering
\begin{equation}
 \frac{\d \mathcal{P}_{\mrm{MI}}}{\d \pT} \rightarrow
 \frac{\d \mathcal{P}_{\mrm{uu}}}{\d \pT} +
 \frac{\d \mathcal{P}_{\mrm{u \delta}}}{\d \pT} +
 \frac{\d \mathcal{P}_{\mrm{\delta u}}}{\d \pT} +
 \frac{\d \mathcal{P}_{\mrm{\delta \delta}}}{\d \pT}
 ,
\end{equation}
where the uu component now represents the original MI probability, the
$\mrm{u \delta}$ and $\mrm{\delta u}$ components a single rescattering and
the $\mrm{\delta \delta}$ component a double rescattering, where both
incoming states to an interaction are previously scattered partons.

\begin{table}
\begin{center}
\begin{tabular}{|l|c|c|c|c|}
 \cline{2-5}
 \multicolumn{1}{c|}{}
 & \multicolumn{2}{|c|}{\textbf{Tevatron}}
 & \multicolumn{2}{|c|}{\textbf{LHC}} \\
 \cline{2-5}
 \multicolumn{1}{c|}{}
 & \textbf{Min Bias} & \textbf{QCD Jets} & \textbf{Min Bias}
 & \textbf{QCD Jets} \\
 \hline
 \textbf{Scatterings}          & 2.81  & 5.11  & 5.21  & 12.20\phantom{0} \\
 \textbf{Single rescatterings} & 0.37  & 1.20  & 0.93  & 3.64             \\
 \textbf{Double rescatterings} & 0.01  & 0.03  & 0.02  & 0.11             \\
 \hline
\end{tabular}
\end{center}
\caption{Average number of scatterings, single rescatterings and double
rescatterings in minimum bias and QCD jet events at Tevatron
($\sqrt{s} = 1.96 \GeV$, QCD jet $\hat{p}_{\perp min} = 20 \GeV$) and LHC
($\sqrt{s} = 14.0 \TeV$, QCD jet $\hat{p}_{\perp min} = 50 \GeV$) energies}
\label{tab:noRes}
\end{table}

\begin{figure}
\begin{center}
\includegraphics[angle=270, scale=1.0]{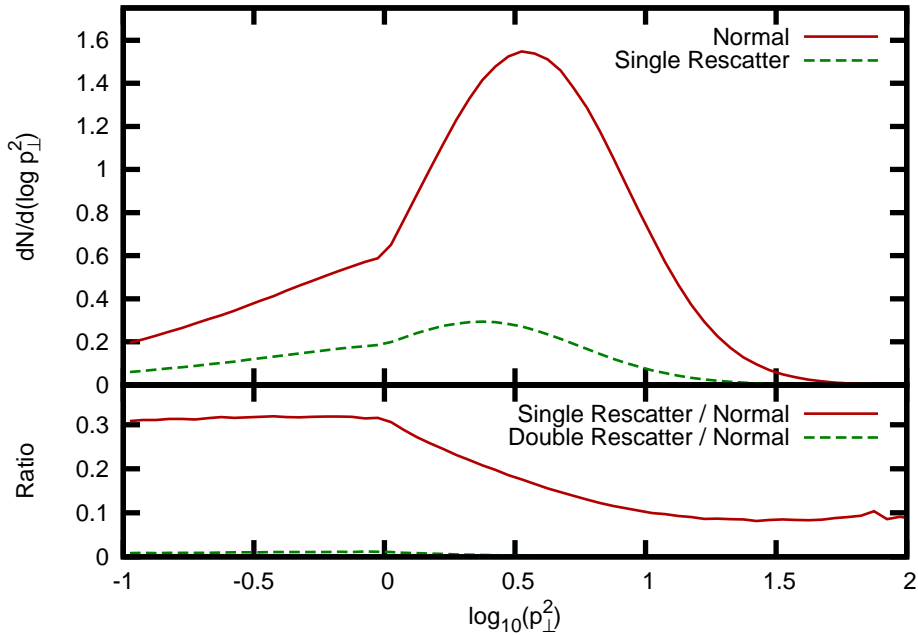}
\end{center}
\caption{Average distribution of scatterings, single rescatterings and
double rescatterings per event ($\sqrt{s} = 1.96 \GeV$, minimum bias).
Double rescattering is not visible at this scale in the
$\d N / \d ( \mrm{log} \, \pT^2)$ plot, but is visible in the ratio}
\label{fig:pTres}
\end{figure}

Some indicative numbers are given in Table~\ref{tab:noRes}, which shows the
average number of scatterings and rescatterings for different types of
event at Tevatron and LHC energies. The average distribution of such
scatterings per event is also shown in Figure~\ref{fig:pTres} for Tevatron
minimum bias events. In the upper plot of
$\d N / \d ( \mrm{log} \, \pT^2)$, the suppression of the cross
section at small $\pT^2$ is caused mainly by the regularisation outlined in
eq.~(\ref{eqn:pt0}), but is also affected by the scaling violation in the
PDFs. Below $\pT^2 \sim 1 \GeV^2$, the PDFs are frozen, giving
rise to an abrupt change in slope. Normal scatterings dominate, but there
is a clear contribution from single rescatterings. In the upper plot, it is
not possible to see the effects of double rescattering, but this is
(barely) visible in the ratio plot below. Given the overall small
contribution from double rescatterings, we neglect these in the following.
As previously predicted, rescattering is a small effect at larger $\pT$
scales, but, when evolving downwards, its relative importance grows as more
and more partons are scattered out of the incoming hadrons and become
available to rescatter. Note that here, we classify the original scattering
and the rescattering by $\pT$, but make no claims on the time ordering of
the two.

\subsection{Mean $\pT$ vs Charged Multiplicity}
While a preliminary framework is in place which allows for hadronic final
states, there are non-trivial recoil kinematics when considering the
combination of rescattering, FSR and primordial $\kT$. With the
dipole-style recoil used in the parton showers, a final-state radiating
parton will usually shuffle momenta with its nearest colour neighbour.
Without rescattering, colour dipoles are not spanned between systems, and
individual systems will locally conserve momentum. With rescattering
enabled, you instead have the possibility of colour dipoles spanning
different scattering systems and therefore the possibility of an individual
system no longer locally conserving momentum. When primordial $\kT$ is now
added through a Lorentz boost, these local momentum imbalances can lead to
global momentum non-conservation. In order to proceed and be able to take
an initial look at the effects of rescattering on colour reconnection, a
temporary solution of deferring FSR until after primordial $\kT$ is added
has been used, as is done in \tsc{Pythia 6.4}.

\begin{figure}
\begin{center}
\includegraphics[angle=270, scale=1.00]{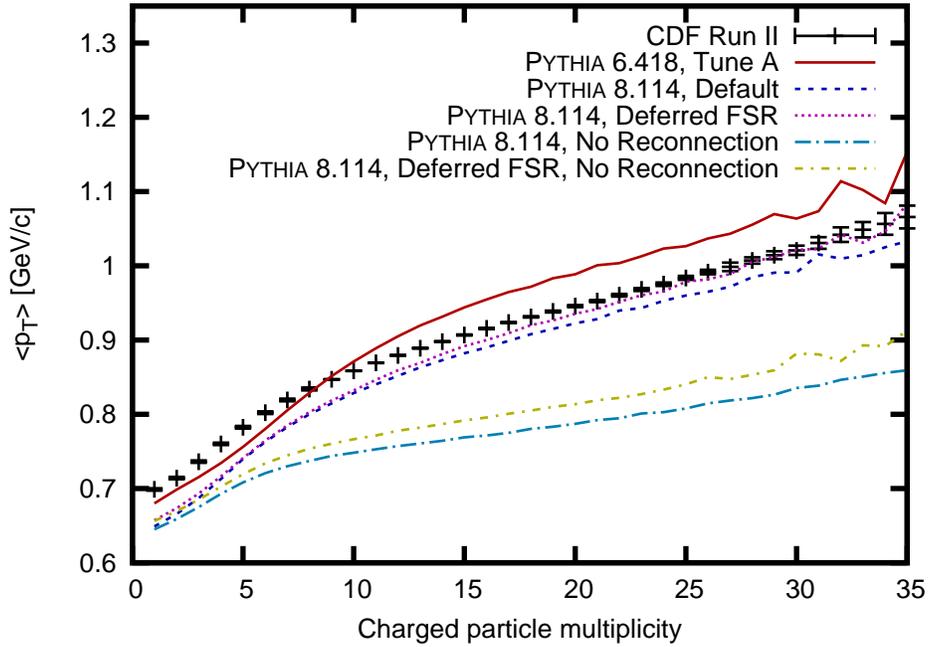}
\end{center}
\caption{Mean $\pT$ vs Charged Multiplicity, $| \eta | \leq 1$ and
$\pT \geq 0.4~\mrm{GeV/c}$, CDF Run II data against Pythia 6.418 (Tune
A) and Pythia 8.114 (default settings) with and without deferred FSR}
\label{fig:meanpt-res-a}
\end{figure}

We begin by studying the mean $\pT$ vs charged multiplicity distribution,
$\langle \pT \rangle(n_{ch})$, from \tsc{Pythia 6.418} (Tune A) and
\tsc{Pythia 8.114} (default settings), compared to the CDF Run II data
($| \eta | \leq 1$ and $\pT \geq 0.4~\mrm{GeV/c}$) \cite{CDFRunII}. For
each run, the ${\pT}_0$ parameter of the MI framework is tuned so that the
mean number of charged particles in the central region is maintained at the
Tune A value.  This is shown in Figure~\ref{fig:meanpt-res-a}, where we can
see that \tsc{Pythia 6}, using virtuality-ordered showers and the old MI
framework, does a reasonable job of describing the data. \tsc{Pythia 8}
does not currently have a full tune to data, but does qualitatively
reproduce the shape of the data when colour reconnection is turned on, up
to an overall normalisation shift. It is clear that without colour
reconnection, the slope of the curve is much too shallow and unlikely to
describe the data, even given an overall shift. The same results with
deferred FSR are also shown; the slope is marginally steeper, but still in
the same region as without deferred FSR.
 
\begin{figure}
\begin{center}
\includegraphics[angle=270, scale=1.00]{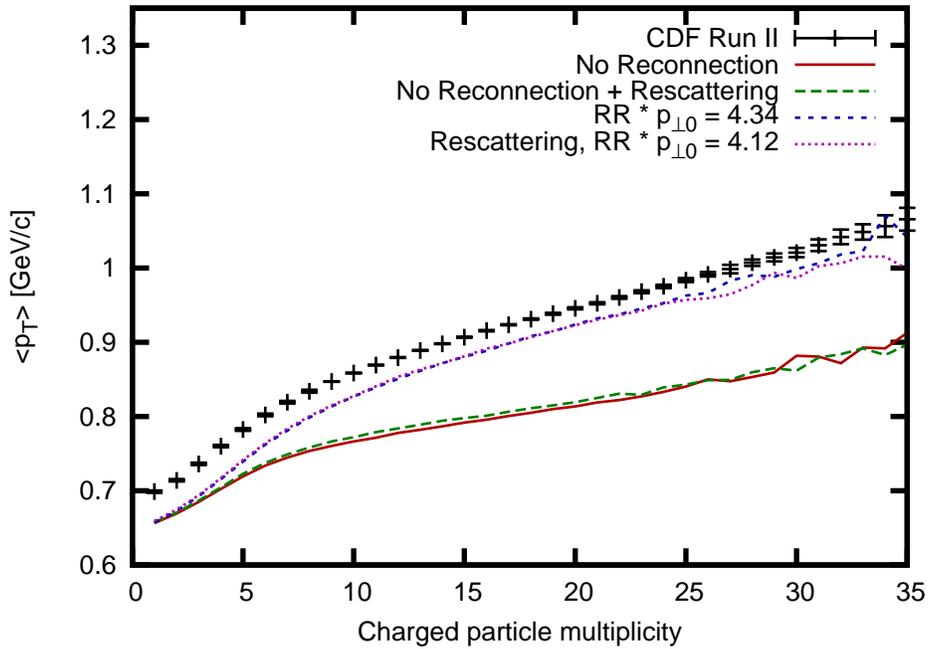}
\end{center}
\caption{Mean $\pT$ vs Charged Multiplicity, \tsc{Pythia 8.114}
(deferred FSR), effects of rescattering}
\label{fig:meanpt-res-b}
\end{figure}

Figure~\ref{fig:meanpt-res-b} now shows the results when rescattering is
enabled. Starting without any colour reconnection, we see that when
rescattering is turned on, there is a rise in the mean $\pT$, but also that
this is in no way a large gain. This is also the case when colour
reconnection is turned on and tuned such that the curve qualitatively
matches the shape of the Run II data. The amount of colour reconnection
used is given in the form $RR * {\pT}_0$, as described in
eq.~(\ref{eqn:crec}). That a rise in the mean $\pT$ is there with
rescattering, but small, is something that was observed already in an early
toy model study. Now, when the full generation framework is almost there,
it is clear that rescattering is not the answer to the colour reconnection
problem. Other potential effects of rescattering remain to be studied.

\section{Enhanced Screening}
\label{sec:ES}
The idea of enhanced screening came from the modelling of initial states
using dipoles in transverse space \cite{Gustafson:2008}. A model using an
extended Mueller dipole formalism has recently been used to describe the
total and diffractive cross sections in $\p\p$ and $\gamma^*\p$ collisions
and the elastic cross section in $\p\p$ scattering \cite{Avsar:2007xg}.
In such a picture, initial-state dipoles are evolved forwards in rapidity,
before two such incoming states are collided. In the model, as the
evolution proceeds, the number of dipoles with small transverse extent
grows faster than that of large dipoles. The dipole size, $r$, determines
the screening length, which appears in the interaction cross section as a
$\pT$ cutoff, ${\pT}_0 \sim 1/r$. Smaller dipoles imply a larger effective
cutoff, and an enhanced amount of screening.  A rough calculation shows
that this screening effect is expected to grow as the square root of the
number of dipoles.

To model this in \tsc{Pythia}, we consider the ${\pT}_0$ parameter of
the MI framework that encapsulates colour screening, as given in
eq.~(\ref{eqn:pt0}). By scaling this value by an amount that grows as the
amount of initial-state activity grows, this enhanced screening effect can
be mimicked. Such a change can be achieved by adjusting the weighting of
the cross section according to
\begin{equation}
 \frac{\d \hat{\sigma}}{\d \pT^2} \propto
 \frac{\alpha_S^2({\pT^2}_0 + \pT^2)}{({\pT^2}_0 + \pT^2)^2}
 \rightarrow
 \frac{\alpha_S^2({\pT^2}_0 + \pT^2)}{(n \, {\pT^2}_0 + \pT^2)^2}
 \label{eqn:es}
 ,
\end{equation}
where $n$ takes a different meaning for two different scenarios. With
the first scenario, ES1, $n$ is set equal to the number of multiple
interactions that have taken place in an event (including the current one).
In the second, ES2, $n$ is set equal to the number of MI+ISR interactions
that have taken place in an event.

\subsection{Mean $\pT$ vs Charged Multiplicity}
\begin{figure}
\begin{center}
\includegraphics[angle=270, scale=1.00]{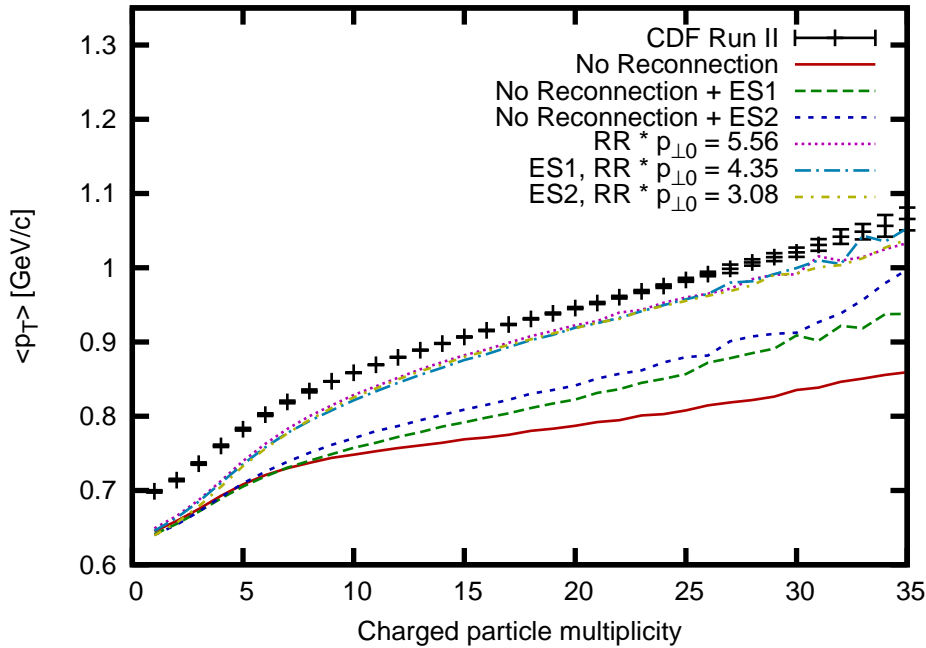}
\end{center}
\caption{Mean $\pT$ vs Charged Multiplicity, \tsc{Pythia 8.114}, effects
of the enhanced screening ansatz}
\label{fig:meanpt-ES}
\end{figure}

We again study the $\langle \pT \rangle(n_{ch})$ distribution, this
time with the enhanced screening ansatz. The results are given in
Figure~\ref{fig:meanpt-ES}. Looking at the curves without colour
reconnection, it is immediately apparent that both scenarios give a
dramatic rise in the mean $\pT$, although not quite enough to explain data
on their own. With colour reconnection now enabled and
tuned, again so that the curves qualitatively match the shape of the Run II
data, it is possible to noticeably reduce the amount of reconnection
needed. With colour reconnection at these levels, there is still perhaps an
uncomfortably large number of systems being reconnected, but the results
are definitely encouraging. There are many more areas to study in relation
to enhanced screening, but from these initial results, it is worth checking
if it may play a role in reducing colour reconnections to a more
comfortable level.

\section{Conclusions}
\label{sec:conclusions}

\tsc{Pythia 8}, the C++ rewrite of the \tsc{Pythia} event generator
has now been released. It has been written with a focus on Tevatron and LHC
applications, something that is evident given the sophisticated MI model
present in the program. The original MI model, introduced in the early
versions of \tsc{Pythia}, has been well proven when compared to
experimental data. The new \tsc{Pythia 8} MI framework, based on this
original model, now generalises the physics processes available, as well as
adding entirely new features.

We have also taken an early look at rescattering and enhanced screening,
two new ideas for modifying the physics inside the MI framework.  There is
currently a preliminary framework for rescattering, although fully
interleaved ISR, FSR and MI is still to come.  It appears, at this early
stage, that rescattering is not the answer to the colour reconnection
problem, but there is still much more to investigate, such as three-jet
multiplicities and other collective effects. The idea of enhanced screening
leads to a simple ansatz that gives large changes when looking at the
$\langle \pT \rangle(n_{ch})$ distribution. Again, there are still many
questions to be asked, including how this modification affects other
distributions.

\end{document}